\documentclass[%
 reprint,
showpacs,
 amsmath,amssymb,
 aps,
]{revtex4-1}

\usepackage{graphicx}
\usepackage{dcolumn}
\usepackage{bm}

\begin{document}

\title{Hyperbolic Blockade: Suppression of the Photonic Density of States and the Spontaneous Emission Rate at the Interface with Conducting Medium}



\author{Evgenii E. Narimanov}
\affiliation{School of Electrical and Computer Engineering  and Birck Nanotechnology 
Center, \\ Purdue University, West Lafayette, IN 47907, USA
}%


\begin{abstract}
Surface scattering of free electrons strongly modifies the electromagnetic response  near the interface.  Due to the inherent anisotropy of the surface scattering that necessarily reverses the normal the interface component of the electron velocity while its tangential component may remain the same, a thin layer near a high-quality interface shows strong dielectric anisotropy. The formation of the resulting hyperbolic dispersion layers  near the metal-dielectric interface strongly modifies the local density of states, and leads to orders of magnitude changes in all associated phenomena.
\end{abstract}

\maketitle


Light incident on a conducting material, changes the dynamics of the free charge carriers near the interface. 
The resulting surface plasmon-polariton excitations  \cite{Maier-book} increase the local photonic density of states, 
leading to a  dramatic change in broad range of related phenomena -- from the  enhancement of 
the spontaneous emission rates near the interface \cite{Surface_Purcell} to surface-enhanced Raman scattering,\cite{surface_Raman}  to subwavelength light localization and confinement \cite{Maier-book}. 
While most of these phenomena can be understood, at least at the qualitative level, within the framework of the 
effective local dielectric permittivity of the metal, this approach becomes progressively more problematic when the plasmon fields change of the scale that is compatible to the electron mean free path. The importance of an accurate
account of the inherent mobility of free charge carriers is now well understood, \cite{Stockman,Mortensen2015,Ciraci2012,PendryHydro}  and
the corresponding ``spatial dispersion'' formalism was successfully used for quantitative description of surface 
plasmon-polaritons in metallic nanostructures. \cite{Stockman,Mortensen2015,Ciraci2012,PendryHydro} 

However, the inherent mobility of the free-charge carriers not only leads to an essentially nonlocal theoretical 
description (the fundamental property which is equally important both at the bulk and near the surface of the conducting medium), but also qualitatively changes the nature of the electromagnetic response near the metal-dielectric interface.
For a high-quality  surface, the electron reflection will reverse normal to the surface component of the momentum, while leaving its tangential projection intact. As a result, while the specular reflection at the interface will not strongly  affect the electromagnetic response in the tangential direction, its  component that is normal to the metal surface, will be substantially altered. Even in the presence of substantial surface roughness,\cite{Ziman} the effect of the surface scattering on the momentum transfer from the free carriers to the  interface (and thus the entire sample as a whole) is 
still very different in the normal and tangential directions. As a result,  the free carrier electromagnetic response  near the conductor - dielectric interface will show strong anisotropy.

In this thin interfacial layer, while diffuse component of the surface scattering leads to an increased loss,  the tangential dielectric permittivity retains its negative sign. However, the electronic contribution to the normal to the interface permittivity is strongly suppressed (as the free-carrier current density at the interface  in this direction is exactly zero, regardless of the 
magnitude of the electric field). As a result, the interface layer has essentially hyperbolic electromagnetic response.

The formation of the hyperbolic layer near the metal-dielectric interface  will no longer support direct resonant coupling from the incident field to the free electrons in the ``bulk'' metal, leading to a suppression of the conventional plasmon 
resonance via the {\it hyperbolic blockade}. While the conventional surface plasmon polariton mode is still present in the system, in can no longer reach the extreme values of the wavenumbers predicted for a ``direct'' (lossy) metal-dielectric interface. At the same time, the hyperbolic layer leads to an additional surface wave -- the so-called ``hyper-plasmon'', that can now co-exist with the standard plasmon-polariton.\cite{ENarxiv} 

As a result, the local photonic density of states (pDOS) at the metal-dielectric interface is strongly modified. First, the peak near the surface plasmon resonance frequency is strongly suppressed, and the corresponding density of states is substantially reduced -- while at other frequencies when the hyper-plasmon waves are present, it can be substantially enhanced. Second, the photonic density of states now shows a very different behavior as a function of the distance to
the metal-dielectric interface $d$. When it's much larger than the thickness of the hyperbolic layer $d_*$, the latter is not ``resolved'' -- and the density of states is close to the value calculated from the  ``bulk'' properties of the metal (albeit with the nonlocal corrections \cite{Stockman}). However, for $d \leq d_*$, it is now the hyperbolic layer that determines the
density of states -- leading to a crossover to a different behavior.

In the quantitative theory of this Letter, we  focus on the calculation of the spontaneous emission rate for a small emitter (such as a dye molecule or a quantum dot) in the proximity to the metal-dielectric interface. While directly connected
to the local density of states via the Fermi Golden Rule, and thus offering a probe into the local pDOS,  the spontaneous emission rate is also an important quantity for both the interpretation of experimental date\cite{Noginov,Vlad,Vinod} and
for technological applications.\cite{PNAS} 

\begin{figure}[htbp] 
   \centering
    \includegraphics[width=3.4 in]{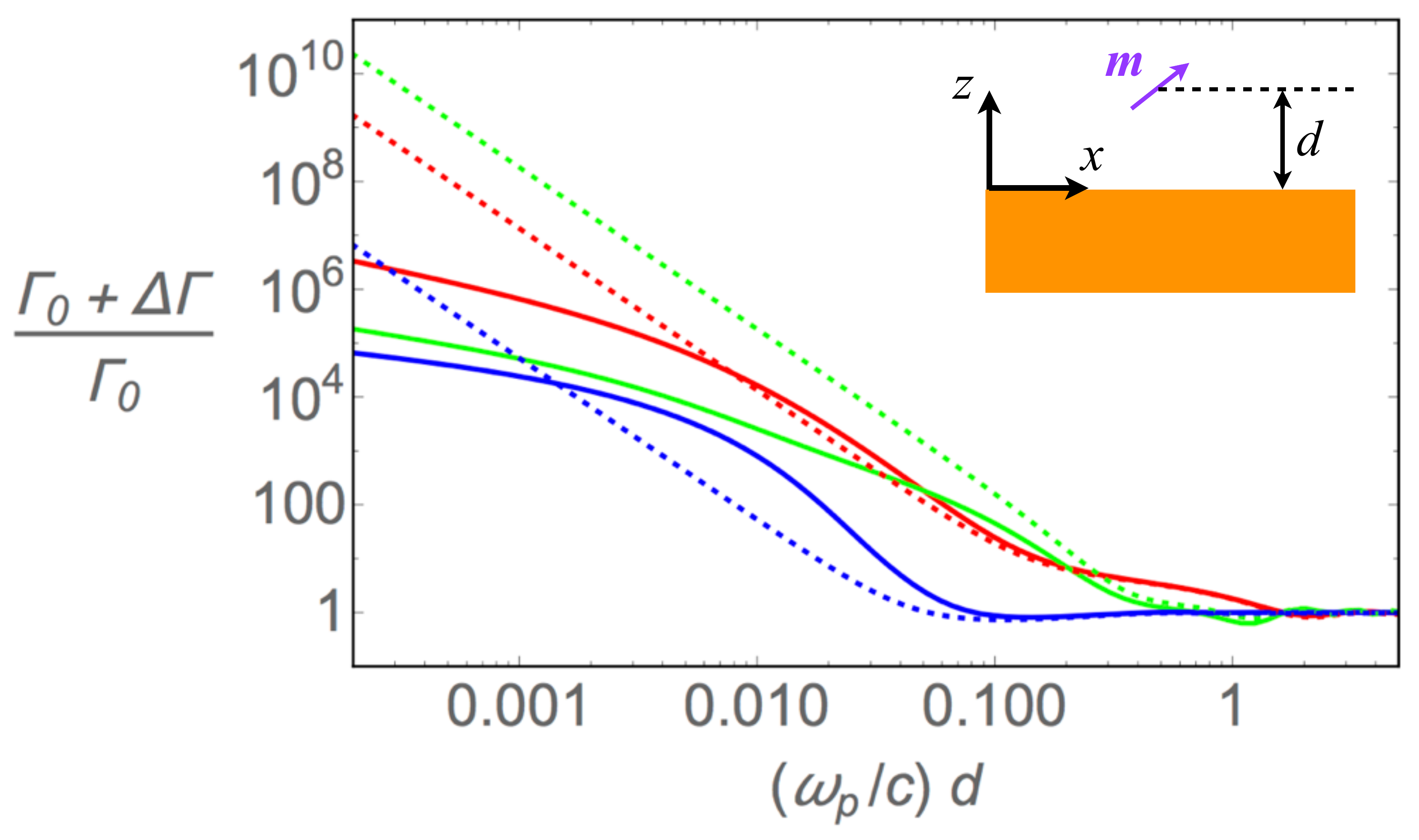}
     \caption{The spontaneous emission rate near the  dielectric - conductor  interface, for the 
 ${\rm AlInAs}/{\rm InGaAs}$ system, \cite{ref:nmat} as a function of the distance $d$ from the emitter to the surface (see
 the inset). The emission rate is normalized to its value in infinite dielectric $\Gamma_0$. Solid lines show the exact solution obtained in the present work, while the corresponding dotted lines represent the results of the calculation based on the local theory. 
 Different colors corresponds to different frequencies and emission polarizations:  dipole moment ${\bf m} \parallel {\bf \hat{n}}$ at $\omega = 0.5 \  \omega_{\rm sp}$ (red), ${\bf m} \perp {\hat{\bf  n}}$   at $\omega = \omega_{\rm sp}$ (green), ${\bf m} \parallel {\hat{\bf  n}}$   at $\omega = 2\  \omega_{\rm sp}$ (blue), where $\omega_{\rm sp}$ is the surface plasmon resonance frequency and $ {\hat{\bf  n}}$ is a unit vector along the normal to the interface. In this calculation, the electron scattering time $\tau = 18.84 / \omega_p$, the crystal lattice permittivity of the conductor $\epsilon_\infty = 12.15$,  the permittivity of the dielectric $\epsilon_d = 10.23$, and the Fermi velocity $v_F = 0.00935$; for the plasma wavelength $\lambda_p \equiv 2 \pi c/\omega_p = 10 \ \mu{\rm m}$ these parameters correspond to the ${\rm AlInAs}/{\rm InGaAs} $ material system of Ref. \cite{ref:nmat}}     
     \label{fig:1}
\end{figure}

In the weak coupling limit,\cite{Surface_Purcell,Leo} for an emitter located at the distance $d$ from a planar interface (see Fig. \ref{fig:1}) we obtain
\begin{eqnarray}
\Gamma & = & \Gamma_0 + \eta \  \Delta\Gamma,
\label{eq:Gamma_Rsp}
\end{eqnarray}
with
\begin{eqnarray}
\Delta \Gamma & = &\frac{3 c^3}{4 \left| {\bf m}\right|^2 \omega^3 \  \epsilon_d^{3/2} } {\rm Re} 
\int_0^\infty \frac{dk \ k}{k_z} \exp\left(2 i k_z d\right) \nonumber \\
& \times & \left[ r_s m_\tau^2 \left(k^2 + k_z^2\right) - r_p \left( m_\tau^2 k_z^2 -
2 m_n^2 k^2 \right)  \right],
\label{eq:dGamma_Rsp}
\end{eqnarray}
where $\eta < 1$ is the quantum efficiency \cite{Leo} that account for other (non-radiative) decay channels of the excited state in the emitter, $k_z \equiv \sqrt{\epsilon_d \left(\omega/c\right)^2 - k^2}$, $\omega$ is the emitted light frequency, and $\epsilon_d$ is the permittivity of the dielectric medium at $z>0$,    while $m_\tau$ and $m_n$ are the tangential and normal to the metal-dielectric interface projections of the unit vector ${\bf m}$ that indicates the direction of the  dipole moment of the emitter (see the inset to Fig. \ref{fig:1}). We emphasize that this expression implies no assumption on the nature of the material on the other side of the interface: the medium can be metallic, hyperbolic or dielectric, with either local or non-local electromagnetic response, as long as it has translational symmetry parallel to the interface, and at least a uniaxial symmetry along the normal to the surface. Under these conditions,  the incident $s-$ and $p-$ polarizations are not mixed up upon reflection, and can be described by the corresponding reflection amplitudes $r_s$ and $r_p$. .

When the distance to the interface $d$   much larger than  de Broglie wavelength of the free charge carriers,
\begin{eqnarray}
d \gg \lambdabar,
\end{eqnarray}
the integral in (\ref{eq:dGamma_Rsp}) is dominated by the waves with in-plane wavenumbers $k \leq 1/d \ll 1/ \lambdabar$. The free charge carrier response at such  wavenumbers can be treated within the semiclassical framework,
via the Boltzmann kinetic equation: \cite{Ziman}
\begin{eqnarray}
\frac{\partial f_{\bf p}}{\partial t}  + {\bf v}_{\bf p} \cdot \nabla f_{\bf p} +
 e {\bf E}\cdot {\bf v}_{\bf p}  \frac{\partial f_0}{\partial \varepsilon_{\bf p}} & = & - \frac{f_{\bf p}  - f_0}{\tau},
\label{eq:Boltzmann}
\end{eqnarray}
where $f_{\bf p}\left({\bf r},t\right)$ is the charge carriers distribution function with its equilibrium (Fermi-Dirac) limit $f_0$, $\varepsilon_{\bf p}$ is the electron energy for the (Bloch) momentum ${\bf p}$,  ${\bf v_p} \equiv \partial\varepsilon_{\bf p}/\partial{\bf p}$ is the corresponding electron group velocity, and $\tau$ is   the effective relaxation time defined by the bulk scattering (due to e.g. phonons, impurities, etc.)

For a high-quality interface along one of the symmetry planes of the crystal, the surface leads to specular reflection of the charge carriers,\cite{Ziman} which can be accounted for by the boundary condition on the distribution function,
\cite{Ziman,Fuchs1938,ReuterSondheimer1948,Sondheimer1950,Soffer1967}
\begin{eqnarray}
f_{{\bf p}^-}\left({\bf r}_s\right) & = &f_{{\bf p}^+}\left({\bf r}_s\right),
\label{eq:ss_specular}
\end{eqnarray}
where  ${bf r}_s$ corresponds to any point at the interface, ${\bf p}^+$ and ${\bf p}^-$ are connected by the specular reflection condition.

The electromagnetic field in the system  defined by the self-consistent solution of the kinetic equation and the surface scattering boundary condition together with the Maxwell equations, where the electron charge and current densities  are given by
\begin{eqnarray}
\rho\left({\bf r}\right) & = & 2 \int \frac{d{\bf p}}{\left(2 \pi \hbar\right)^3} \cdot  \left(f_{\bf p}\left({\bf r}\right)  -f_0\left(\varepsilon_{\bf p} \right) \right) , \label{eqn:rho} \\
{\bf j}\left({\bf r}\right) & = & 2 \int \frac{d{\bf p}}{\left(2 \pi \hbar\right)^3} \cdot  e \  {\bf v}_{\bf p} f_{\bf p}\left({\bf r}\right),   \label{eqn:current}
\end{eqnarray}

Following the mathematical  approach described in Ref. \cite{ENarxiv}, this problem can be solved exactly, and for the reflection coefficient in the $s$-polarization we obtain
\begin{eqnarray}
r_s & = & - 1 + {2 k_z }\left(k_z + \sqrt{ \epsilon_\tau\left(k\right)\cdot \left(\omega/c\right)^2 - k^2} \ \right)^{-1},
\label{eq:rs}
\end{eqnarray}
where
\begin{eqnarray}
\epsilon_\tau\left(k\right) & = & \epsilon_\infty  + \frac{2 e^2}{\pi^2 \hbar^3 \omega}
 \int\frac{d{\bf p} \ v_y^2 }{ \omega  +  k  v_y  + i/\tau }  \cdot  \frac{\partial f_0}{\partial \varepsilon_{\bf p}}.  \label{eq:et} 
\end{eqnarray}
For a degenerate electron gas the integration in (\ref{eq:et}) yields
\begin{eqnarray}
\epsilon_\tau\left(k\right) & = & \epsilon_\infty - \frac{\epsilon_\infty  \ \omega_p^2}{\omega \left(\omega + i / \tau\right)} \  {\cal F}_\tau\left(\frac{v_F  k}{\omega + i/\tau}\right),
\label{eq:etd}
\end{eqnarray}
with
\begin{eqnarray}
{\cal F}_\tau\left(x \right) & = & \frac{3}{x^2} \left( {\cal F}_0\left(x\right) - 1\right) ,
\ \ 
{\cal F}_{0}\left(x\right)  =  \frac{1}{2 x}
\log\frac{1+x}{1-x}, \ \ \ 
\end{eqnarray}
where   $v_F$ is the electron Fermi velocity, $\epsilon_\infty$ is  the ``background'' permittivity of the crystal lattice in the conductor, and $\omega_p$ is the standard plasma frequency.\cite{Maier-book} Note that the expression $\epsilon_\tau(k)$ in Eqn. (\ref{eq:etd}) is consistent with the other models of nonlocal free carriers response used in the recent literature. \cite{Stockman,Khurgin}

For the $p$-polarization, we find
\begin{eqnarray}
r_p & = & - 1 + {2 k_z}\left( k_z +   \frac{2 i \epsilon_d \omega^2}{\pi c^2}  \int_{0}^\infty 
\frac{dq}{D\left(k, q\right)} \right)^{-1},
\label{eq:rp} 
\end{eqnarray}
where
\begin{eqnarray}
D\left(k,q\right) & = & \epsilon_{x}\left(k, q\right) \frac{\omega^2}{c^2} - q^2   -  \frac{\nu^2_{xz}\left(k,q\right)}{\epsilon_{z}\left(k, q\right) \frac{\omega^2}{c^2} - k^2 }, 
\label{Eq:D_SM}
\end{eqnarray}
and
\begin{eqnarray}
 \epsilon_{x,z}\left(k,q\right) & = &\epsilon_\infty  - \frac{16 \pi i e^2 \tau}{\omega}
 \int_{v_z > 0}\frac{d{\bf p}}{\left(2 \pi \hbar\right)^3}   \frac{\partial f_0}{\partial \varepsilon_{\bf p}}
 \nonumber \\
 & \times & v_{x,z}^2 \ \frac{ 1 - i \omega \tau + i k  v_x \tau  }{ \left( 1 - i \omega \tau + i k  v_x \right)^2 +q^2 v_z^2 \tau^2 },
 \label{eq:exz}
\\
 \nu_{xz}\left(k,q\right) & = &k q -  \frac{16 \pi e^2 \tau^2 \omega q}{c^2}  
 \int_{v_z > 0}\frac{d{\bf p}}{\left(2 \pi \hbar\right)^3}  \frac{\partial f_0}{\partial \varepsilon_{\bf p}}  \nonumber \\
& \times & v_x v_z^2 \ \frac{  1 }{ \left( 1 - i \omega \tau + i k  v_x \tau \right)^2 +q^2 v_z^2 \tau^2 }. \ \ \  \label{eq:nxz}
\end{eqnarray}
For a degenerate electron gas, \cite{footnote3} analytical integration over the electron momentum ${\bf p}$ reduces Eqns. (\ref{eq:exz}),(\ref{eq:nxz}) to
\begin{eqnarray}
&  & \epsilon_{x} \left(k,q\right)   =   
\epsilon_\infty - \frac{3 \epsilon_\infty}{2} 
\frac{\omega_p^2}{v_F^2} \frac{1 + i / \left(\omega\tau\right) }{k^2 + q^2} 
\left\{   
\frac{q^2 - 2 k^2}{k^2 + q^2}   \right. \nonumber \\
&   + &    \left.  \left(\frac{v_F^2 q^2}{\left(\omega + i/\tau\right)^2 } + \frac{2 k^2 - q^2}{k^2 + q^2} \right)  {\cal F}_0\left(\frac{v_F \sqrt{k^2 + q^2}}{\omega + i/\tau}\right)
 \right\},  \ \  \ \ \ \ \label{eq:exd}
\\
 & & \epsilon_{z}\left(k,q\right)  =  
\epsilon_{x}\left(q, k\right), \\
& & \frac{\nu_{xz}\left(k,q\right)}{k q }   = 
1 +  \frac{9 \epsilon_\infty}{2 \left(1 + \frac{i}{\omega \tau}\right) } \frac{{\cal F}_{\nu}\left(\frac{v_F \sqrt{k^2 + q^2}}{\omega + i/\tau}\right)}{\left(k^2 + q^2\right) c^2 / \omega_p^2 },
\end{eqnarray}
where
\begin{eqnarray}
{\cal F}_{\nu}\left(x\right) & = & \frac{1}{x^2} + \left( \frac{1}{3} - \frac{1}{x^2}\right)
{\cal F}_0\left(x\right). \label{eq:Fnu}
\end{eqnarray}
Together, Eqns. (\ref{eq:rp}) and (\ref{eq:exd})-(\ref{eq:Fnu}) define the reflection coefficient $r_p$.

The resulting spontaneous emission rate can be calculated by substituting our analytical expressions for the reflection coefficients $r_s$ and $r_p$ into the general equation (\ref{eq:dGamma_Rsp}). In Fig. \ref{fig:1} we compare the resulting values (solid lines) with the predictions of the  standard local theory (dashed lines) that describes the conductor as an effective medium with the (Drude) permittivity $
\epsilon_m\left(\omega\right)  =  \epsilon_\infty \left(1 - \frac{\omega_p^2}{\omega \left(\omega + i/\tau\right)} \right)$. Note that, as the distance to the interface is reduced, local approximation initially underestimates the density of states. This is consistent with the results of the existing non-local theories. \cite{Stockman}  However, at a smaller distance $d < d_*$, this behavior is reversed: the actual density of states is now smaller than the local estimate. This is the result of the hyperbolic blockade introduced in the present work: the hyperbolic layer ``blocks'' the coupling to conventional surface plasmon-polaritons, and the photonic density of states is reduced. Also note strong frequency dependence of $d_*$:  the distance corresponding to the cross-over between the two different regimes, non-monotonically changes with the electromagnetic wavelength. 

\begin{figure}[htbp] 
   \centering
    \includegraphics[width=3.3in]{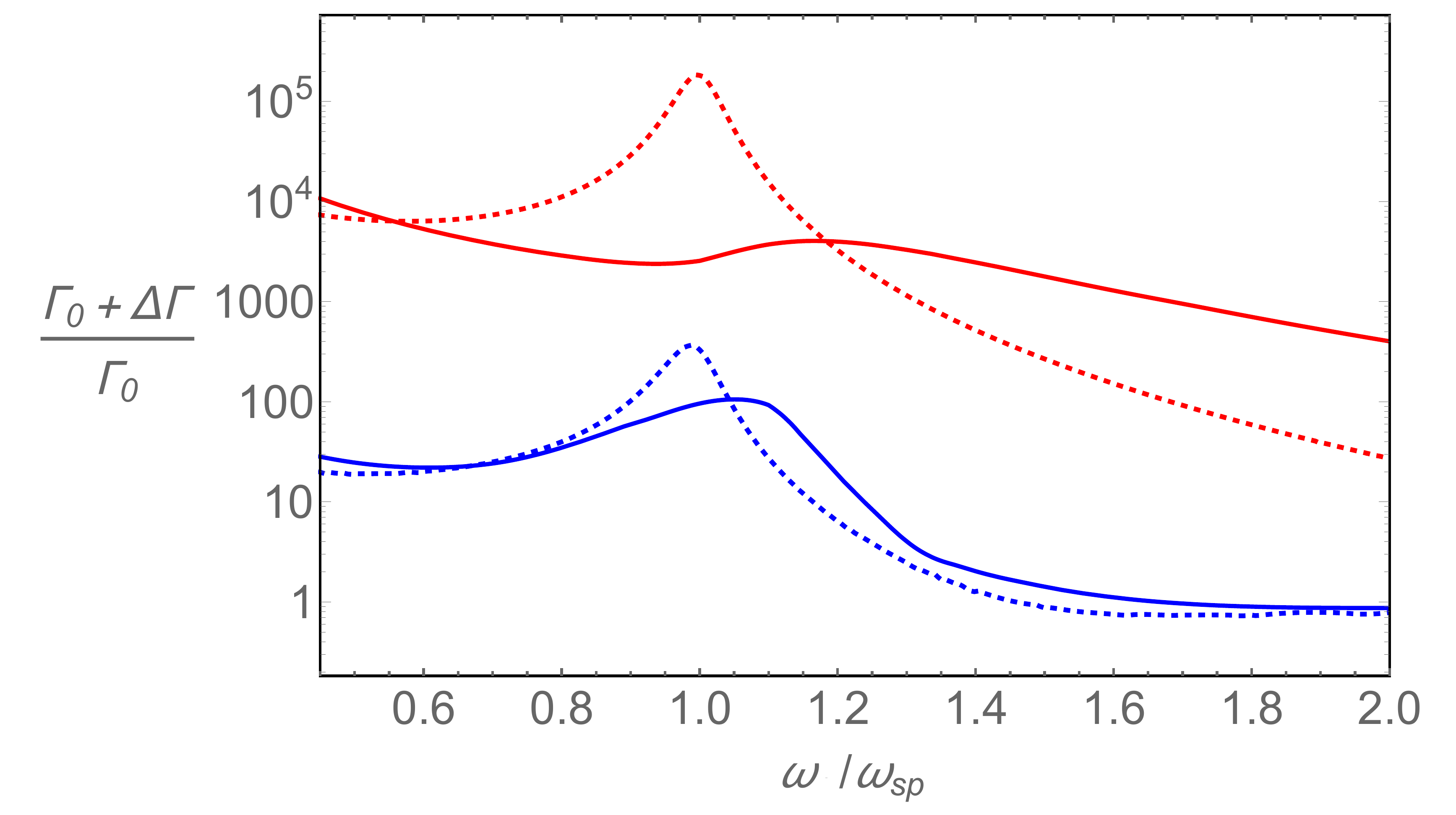}
       \caption{The frequency dependence of the spontaneous emission rate near the conductor-dielectric interface. As in Fig. \ref{fig:1},  solid lines show  the exact solution, while the dotted curves correspond to the calculations using the local response model, for ${\bf m} \parallel {\bf \hat{n}}$ at $d = 0.01 c/\omega_p$ (red) and $d = 0.1 c/\omega_p$ (blue). The  material parameters are the same as in Fig. \ref{fig:1}. Note the suppression of the plasmon resonance due to the hyperbolic blockage, together with and order of magnitude the enhancement of the spontaneous emission rate above the plasmon resonance frequency seen at smaller distance to the interface. 
     }
          \label{fig:2}
\end{figure}

The frequency dependence of the spontaneous emission rate for a given distance to the interface, presented in Fig. \ref{fig:2}, shows further evidence of the hyperbolic blockade. Note the suppression of the plasmon resonance, especially at the smaller distance to the interface. Furthermore, the coupling to hyper-plasmons -- the new surface waves that originate from the hyperbolic layer, \cite{ENarxiv}  manifests itself in the enhancement of the spontaneous emission rate, seen in Fig. \ref{fig:2}  at higher frequencies. 

\begin{figure}[htbp] 
   \centering
    \includegraphics[width=3.25in]{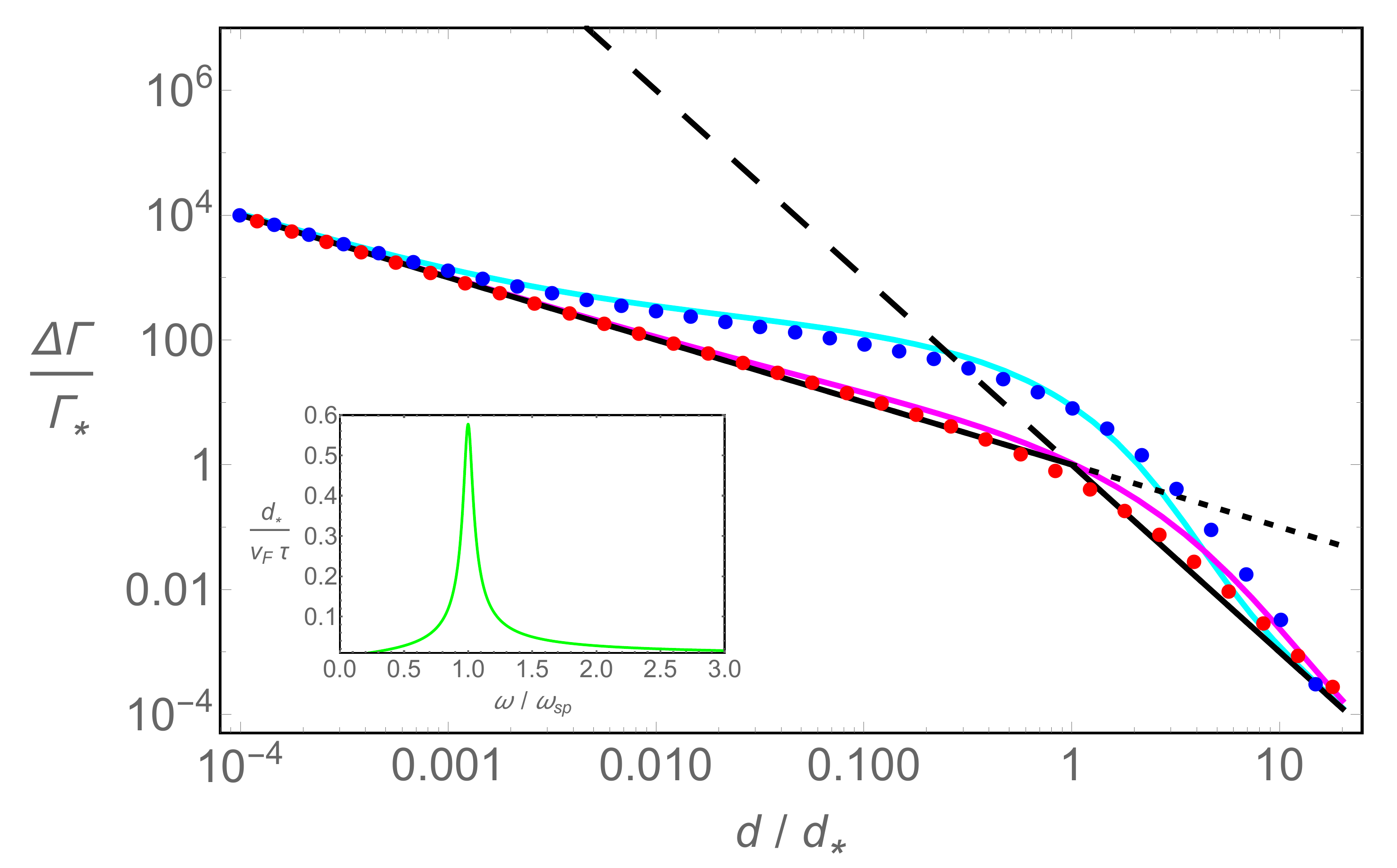}
     \caption{The spontaneous emission rate as a function of the distance to the interface, in scaled coordinates. Red dots and the red curve correspond to the exact solution and the approximation of Eqn. (\ref{eq:GA1}) for $\omega = 0.5 \ \omega_{\rm sp}$, while blue dots and the blue curve show the exact solution and the approximation of Eqn. (\ref{eq:GA1}) for $\omega = 2 \ \omega_{\rm sp}$. The dipole moment ${\bf m} \parallel {\bf \hat{n}}$. Solid black line corresponds to the interpolation (\ref{eq:GA}), with the black dotted and dashed lines indicating  the $ d \ll d_*$ and $d_* \ll d \ll \lambda_0$ limits of the exact solution. The inset shows the frequency variation of $d_*$.}     
     \label{fig:3}
\end{figure}

When the distance from the emitter to the interface is much smaller than the free-space wavelength, $d \ll \lambda_0$,
the analytical expression for the spontaneous emission rate can be reduced to
 \begin{eqnarray}
\Delta \Gamma & = & \frac{3 }{4 }  \ \Gamma_0\ \frac{m_\tau^2 + 2 \ m_n^2}{\left| {\bf m} \right|^2}   \   \frac{ \epsilon_\infty}{\sqrt{\epsilon_d} \left(\epsilon_d + \epsilon_\infty\right)^2} \left(\frac{c}{v_F}\right)^2 \left( \frac{\omega_p}{\omega} \right)^2
\nonumber \\
& \times &\left\{ 
\frac{3}{2 \omega \tau}  \frac{ c}{  \omega  d} 
+  \frac{c}{v_F}  \frac{ \epsilon_d + \epsilon_\infty}{\epsilon_d  +\epsilon_m\left(\omega\right)} \ 
  {\rm Im} \left[  \left(1 + \frac{i}{ \omega  \tau} \right)^2 \right. 
\right. 
\nonumber \\
& \times & \left.  \left.   \sum_{\alpha = 1}^3  
\frac{u_\alpha^5}{\Pi_{\beta \neq \alpha}  \left(u_\alpha -u_\beta \right)} {\cal Q}\left(\frac{2 \left(\omega + i/\tau\right) d}{v_F} u_\alpha  \right) \right]
 \right\}, \ \  \ \  \ \ \label{eq:GA1}
\end{eqnarray}
where ${\cal Q}$ is related to the incomplete gamma-funciton of $0$-th order
\begin{eqnarray}
{\cal Q}\left(x\right) & = & \exp\left(- x\right) \Gamma\left(0, - x\right),
\label{eq:GQ}
\end{eqnarray}
and
\begin{eqnarray}
u_1  = \zeta\left(\mu\right) + \frac{\mu}{\zeta\left(\mu\right)},  \ \ 
u_{2,3} & = &  e^{\pm \frac{2 i \pi}{3}} \zeta\left(\mu\right) +  \frac{e^{\mp \frac{2 i \pi}{3}} \mu}{\zeta\left(\mu\right)}, \ \ \ \ \ \label{eq:u123} 
\end{eqnarray}
with
\begin{eqnarray}
\zeta\left(\mu\right)  =  \sqrt[3]{2 i \mu + \sqrt{-\mu^3 - 4 \mu^2}}, \ \ \ \mu  =  \frac{1}{2} \frac{\epsilon_d + \epsilon_m\left(\omega\right)}{\epsilon_d + \epsilon_\infty } \ \ \ \ \ 
\end{eqnarray}
In Fig. \ref{fig:3}, we compare the predictions of Eqn. (\ref{eq:GA1}) (colored lines)  with the corresponding results of 
the exact calculations (colored dots), as functions of the distance to the interface, for two different frequencies. Note excellent agreement in the entire parameter range shown in the figure. 

Depending on the relative value of the distance $d$ and the ``electronic'' scale
$\ell \equiv {v_F} \cdot {\rm min}\left[\tau, 1/{\omega} \right]$, the analytical expression (\ref{eq:GQ}) has the limiting behavior
\begin{eqnarray}
\frac{\Delta \Gamma}{\Gamma_0}  & = & \frac{m_\tau^2 + 2 \ m_n^2}{2 \left| { \bf m} \right|^2}  
\left\{
\begin{array}{cc}
\gamma_0\left(\omega, d \right), & d \ll  \ell \\ 
\gamma_\infty\left(\omega,d  \right), & \ell \ll  d \ll \lambda_0
\end{array}
\right.  ,
\label{eq:GA2}
\end{eqnarray}

\noindent
where
\begin{eqnarray}
\gamma_0\left(\omega, d \right) & = &  \frac{9}{4} \frac{\epsilon_\infty \ \omega_p^2}{\left(\epsilon_d + \epsilon_\infty\right)^2 \omega^3 \tau} \left(\frac{c}{v_F}\right)^2 \frac{c}{\sqrt{\epsilon_d} \omega d}, 
\label{eq:g0}
\end{eqnarray}
and
\begin{eqnarray}
\gamma_\infty\left(\omega, d \right) & = &  \frac{3}{8} \ {\rm Im}\left[ 
\frac{\epsilon_m\left(\omega\right) -  \epsilon_d }{\epsilon_m\left(\omega\right) + \epsilon_d} \right] \left( \frac{c}{\sqrt{\epsilon_d}  \omega d}\right)^3.
\label{eq:ginf}   
\end{eqnarray}
Eqn. (\ref{eq:GA1}) can therefore be further approximated by the interpolating function 
\begin{eqnarray}
\Delta \Gamma\left(d\right)  & = & \Gamma_* \cdot
\left\{
\begin{array}{lc}
\  {d_*}/{d}, & d \leq  d_* \\ 
\left({d_*}/{d}\right)^3, & d \geq  d_*
\end{array}
\right.,
\label{eq:GA}
\end{eqnarray}
where
\begin{eqnarray}
\frac{\Gamma_*}{ \Gamma_0} & = & \frac{m_\tau^2 + 2 \ m_n^2}{2 \left| { \bf m} \right|^2} 
\frac{ 9 \left(c/v_F\right)^2 \omega_{\rm sp}^2}{4 \left(\epsilon_d + \epsilon_\infty\right)\omega^3 \tau}
\frac{c}{\sqrt{\epsilon_d} \omega d_*},
\end{eqnarray}
and
\begin{eqnarray}
d_* & = & \frac{1}{\sqrt{3}} \frac{v_F \tau}{\sqrt{1 + \left(\omega\tau\right)^2 \left(1
- \omega_{\rm sp}^2 / \omega^2 \right)^2}}.
\end{eqnarray}
Here, $\omega_{\rm sp}$ is described by  the standard expression for the frequency of the surface plasmon resonance at the planar interface of a dielectric with Drude metal, $\omega_{\rm sp} = {\omega_p}/ {\sqrt{1 + \epsilon_d / \epsilon_\infty}}$.

The solid black line in Fig. \ref{fig:3} plots Eqn. (\ref{eq:GA}), while the  dashed and dotted lines correspond to $\gamma_\infty\left(\omega, d \right)$ and  $\gamma_0\left(\omega,d  \right)$ respectively. Although not sufficiently accurate at the quantitative level, the interpolation (\ref{eq:GA}) correctly represents the qualitative behavior of the spontaneous emission rate and adequately describes the cross-over between the two regimes.

The inset of Fig. \ref{fig:3} shows the frequency dependence of the hyperbolic layer thickness $d_*$. Note its non-monotonic behavior, noticed earlier in the context of the general behavior of the spontaneous emission rate as a function of the distance to the interface (see Fig. \ref{fig:1}). The regime $d < d_*$ corresponds to the suppression of the 
plasmon resonance due to the hyperbolic blockade. Except for $\omega = \omega_{\rm sp}$ when $d_* \sim v_F \tau $, as a  function of frequency $d_*(\omega)$ behaves as $v_F /\omega$ at $\omega > \omega_p$ and as $v_F \omega / \omega_p^2$ for $\omega < \omega_p$, with the characteristic scale given by the Thomas-Fermi screening length $\sim v_F / \omega_p$. For a good metal, in the optical  range $d_*$ is on the order of a nanometer, which makes the limit $d \leq d_*$ essentially inaccessible. On the other hand, in transparent conducting oxides such as the ITO \cite{Sasha_review} or in doped semiconductors,\cite{ref:nmat,Wasserman2015}  we find $d_*$ on the order of a few tens of nanometers -- and the regime $d < d_*$ corresponds to the common situation of an active quantum well in a close proximity to a doped semiconductor substrate. In this case, the phenomenon of the hyperbolic blockade and the theory introduced in the present work, are essential for the accurate account of light emission from such systems.

This work was partially supported by the National Science Foundation (grant 1629276-DMR), Army Research Office (grant W911NF-14-1-0639) and Gordon and Betty Moore Foundation.


\begin{thebibliography}{10}
\bibitem{footnote3}The inequality $\varepsilon_F \gg k_B T$  is generally well satisfied for plasmonic materials even at room temperature: while $k_B T \simeq 20 \ {\rm meV}$, the Fermi energy is usually on the order of an electron-volt.
\end{thebibliography}
\end{document}